\documentclass[twocolumn,preprintnumbers,showpacs,amsmath,amssymb]{revtex4}

\usepackage{psfrag}
\usepackage{graphicx}% Include figure files
\usepackage{dcolumn}% Align table columns on decimal point
\usepackage{bm}% bold math
\usepackage{epsfig}
\usepackage{color}

\begin{document}

\title{Back-reaction of perturbation wave packets on gray solitons}% Force line breaks with \\

\author{P.B. Walczak}

\author{J.R. Anglin}
\affiliation{%
Physics Department and State Research Center OPTIMAS, University of Kaiserslautern,
Erwin-Schr\"odinger-Str. 46, D-67663 Kaiserslautern, Germany
}%

\date{\today}

\begin{abstract}
Within the Bogoliubov--de Gennes linearization theory of quantum or classical perturbations around a background solution to the one-dimensional nonlinear Schr\"odinger equation, we study the back-reaction of wave packet perturbations on a gray soliton background. From our recently published exact solutions, we determine that a wave packet effectively jumps ahead as it passes through a soliton, emerging with a wavelength-dependent forward translation in comparison to its motion in absence of the soliton. From this and from the full theory's exact momentum conservation, we deduce that post-Bogoliubov back-reaction must include a commensurate forward advance by the soliton itself. We quantify this effect with a simple theory, and confirm that it agrees with full numerical solution of the classical nonlinear Schr\"odinger equation. We briefly discuss the implications of this effect for quantum behavior of solitons in quasi-condensed dilute gases at finite temperature.

\end{abstract}

\pacs{03.75.Lm, 03.65.Ge, 03.75.Kk}
\maketitle

%%%%%%%%%%%%%%%%%%%%%%%%%%%%%%%%%%%%%%%%%%%%%%
%
%%%%%%%%%%%%%%%%%%%%%%%%%%%%%%%%%%%%%%%%%%%%%%
\section{Introduction}
\label{intro}

The linear evolution of perturbations around known exact solutions is one of the most basic tools of theoretical physics. It is invoked constantly in almost every field. While the perturbative approach is conceptually obvious, this does not make it naive or trivial. It is typically well justified as the first term in a systematic expansion in a small parameter of the full theory, and it can be very non-trivial indeed, in cases where the background solution being perturbed is itself non-trivial. Hawking radiation \cite{Hawking1974,Hawking1975,Unruh1976}, as derived for linearized quantum fields propagating in fixed spacetime metrics with event horizons, is perhaps the most exotic example of this very common general scenario. From the very fact that the perturbations can be so dramatically affected by the background configuration, however, one must expect on the basis of Newton's Third Law of Motion that the perturbations should in turn evoke changes in background itself. This is known as \textit{back-reaction}. 

In quantitative and technical terms, back-reaction is simply one component of the continuing perturbative expansion beyond leading order; but it is often qualitatively and conceptually much more significant than any small, higher-order corrections to the evolution of the rest of the excitation modes. The paradigm of perturbations around a fixed background is so important that it often determines our entire conception of a physical system. The fact that the background is not fixed may be merely a higher order perturbation in a mathematical sense, but it may often induce a non-perturbatively large change in how we think about the system as a whole.

Because of their extreme controllability, Bose-Einstein condensates (BEC) of atomic gases \cite{Weller2008,Stellmer2008,Becker2008} provide an excellent test system in which to study back-reaction as a fundamental problem. Quasi-one-dimensional condensates can serve especially well in this way, because general solutions to the linearized problem may be available for them in explicit closed form, even for non-trivial backgrounds. One class of such non-trivial backgrounds in one dimension are the gray soliton solutions to the nonlinear Schr\"odinger equation (NLSE) with repulsive nonlinearity \cite{Tsuzuki1971}, which is known in its role as the mean field theory for a dilute Bose gas as the Gross-Pitaevskii equation (GPE) \cite{Gross1961,Pitaevskii1961}. We have recently presented a complete set of exact solutions to the linearization of the NLSE (known in the many-body context as the Bogoliubov-de Gennes equation) around any gray soliton background \cite{Walczak2011}. This provides a rigorous basis on which to examine the back-reaction effects of perturbations upon solitons.

As a qualitative fact it is well recognized that solitons are indeed affected by perturbations, including quantum and thermal fluctuations. The integrability of the one-dimensional NLSE with no spatial inhomogeneity makes the interaction between solitons and other excitations rather special, but even in this limit it is not entirely eliminated. In cases where integrability is broken, for instance by a trapping potential, interactions between solitons and perturbations are of course more general. The interaction between solitons and general time-dependent long wavelength hydrodynamic excitations was derived in \cite{Busch2000} according to a multiple scale treatment within the classical one-dimensional GPE with slowly varying trapping potentials. Dissipation of solitons via scattering with excitations when three-dimensional effects break integrability were discussed in \cite{Muryshev2002}, and more recently in \cite{Mazets2008}. In \cite{Martin2010,Martin2010a} soliton evolution was analyzed within the so-called truncated Wigner approximation \cite{Steel1998}, in which the initial condensate state is sampled according to quantum statistics, but its dynamics is treated classically. A different mean-field-based approach was used in \cite{Cockburn2010}, based on a stochastic and dissipative Gross-Pitaevskii equation (GPE) \cite{Stoof1999} derived as an approximate treatment for the full many-body dynamics. Still other recent work towards identifying the effect on solitons of their embedding in the larger dynamical system of the quasi-condensate has numerically solved a fully quantum discretized version of the problem, computing correlation functions from which some conclusions about soliton behavior may be drawn \cite{Mishmash2009,Mishmash2009a}. 

The contribution of this present paper is complementary to all of the above previous work. On the one hand we treat only the integrable homogeneous problem and reach one perturbative order beyond the linear. On the other hand, however, our results are analytical and explicit, and they are based on exact solutions to the linearized problem combined with the exact constraints of momentum and number conservation.

In this paper we analytically study the leading order dynamical back-reaction of traveling small-amplitude perturbations on one-dimensional gray solitons. Based on the analytical solutions to the Bogoliubov--de Gennes equation (BdGE) \cite{Walczak2011}, we first show that in the presence of a gray soliton elementary excitations experience a well-defined phase shift within Bogoliubov approximation, while the soliton is unaffected. Wave packets thereby undergo a position shift when passing the soliton. By including post-Bogoliubov terms and applying momentum and number conservation, we find a corresponding position shift of the soliton, and derive an analytical formula for this deterministic back-reaction effect. Comparison with numerical integration of the GPE verifies our analytical results classically. Our linear results are just as valid quantum mechanically as they are classically, and the quantum version of our post-Bogoliubov calculations introduces only a trivial operator ordering question. Since the conservation laws on which our conclusions are based remain valid quantum mechanically, therefore, our analysis can be straightforwardly extended to the quantum problem.

The paper is organized as follows. In Sec.~\ref{soliton} the gray soliton as a class of solutions to the GPE and and the elementary excitations around the gray-soliton background as solutions to the BdGE are reviewed, with special focus on their asymptotic behavior away from the soliton. In Sec.~\ref{creation} we construct excitation pulses of finite wave length in the gray-soliton background by superposition of the elementary excitations and find a displacement of the pulses when crossing the soliton. In Sec.~\ref{secondorderprocess} we derive an analytical formula for the back-reaction of the soliton due to the excitations, which needs inclusion of post-Bogoliubov terms, and compare it with numerical time propagation under the GPE. In Sec.~\ref{discuss} we summarize our results and briefly discuss some implications of our results.

%%%%%%%%%%%%%%%%%%%%%%%%%%%%%%%%%%%%%%%%%%%%%%
% soliton background + general things
%%%%%%%%%%%%%%%%%%%%%%%%%%%%%%%%%%%%%%%%%%%%%%
\section{Gray-soliton background and elementary excitations}
\label{soliton}
We begin by discussing gray solitons as background field configurations \cite{Tsuzuki1971,Busch2000}. These are a family of mean field wave functions parametrized by the soliton position $x_0$, constant soliton velocity $\beta$, and fixed background velocity $v$ and asymptotic density $\mu = c^{2}$:
\begin{equation}
\label{sol}
\begin{split}
\psi_0(x) &=e^{-\mathrm{i} v (x-x_0)} \{ \mathrm{i} \beta + \kappa \tanh \kappa (x-x_0) \}\\
\kappa &=\sqrt{\mu -\beta^2}\;.
\end{split}
\end{equation}
The constant $\kappa$ must be real, and consequently the soliton speed $ \beta$ cannot exceed the speed of sound $c=\sqrt{\mu}$ of the background field. For increasing $\beta$ the density depression around the soliton position $x_0$ becomes shallower until the gray soliton reduces to a uniform background for $\beta \to \pm \sqrt{\mu}$. The gray soliton can be considered a localized object, since the local density $|\psi_{0}|^{2}$ approaches the asymptotic value $\mu$ exponentially rapidly for $\kappa |x-x_0|\gg 2$. The phase of $\psi_{0}$ likewise rotates within the localized region $\kappa |x-x_0|\lesssim 2$ between asymptotically constant values. With a compatible choice of the gas velocity $v$, therefore, the soliton \eqref{sol} also represents a suitably periodic wave function for a finite ring-like system of perimeter $2L$, up to negligible errors of order $ \mathcal{O}(e^{- 2 \kappa L})$ \cite{Walczak2011}. 

The soliton background \eqref{sol} is an exact solution to the NLSE/GPE in $1+1$ dimensions (one dimension of space and one of time) \cite{Pitaevskii2003}. For repulsive point-like interaction among the particles and uniform potential this reads
\begin{equation}
\label{GP1}
\mathrm{i} \partial_{t} \psi(x, t) = \left( -\frac{1}{2} \partial_{x}^2 + \mathrm{i} (\beta-v) \partial_{x} + |\psi|^2 - \tilde{\mu} \right) \psi \;,
\end{equation}
expressed in appropriate dimensionless variables and in a frame moving with velocity $\beta-v$ relative to the laboratory frame. We have defined
\begin{equation}
\label{chemical}
\tilde{\mu} \equiv \mu+v \beta-\frac{v^2}{2} \; ,
\end{equation}
to shorten many equations. Thus, \eqref{sol} represents the gray soliton in a frame co-moving with the soliton. In the context of Bose gases the NLSE \eqref{GP1} is obtained from the Heisenberg equation of motion for the second-quantized particle destruction operator field in mean field approximation. Since the soliton position $x_{0}$ is kept constant by our adoption of the co-moving frame, we hereafter set $x_0 = 0$ without loss of generality.

%%%%%%%%%%%%%%%%%%%%%%%%%%%%%%%%%%%%%%%%%%%%%%
\subsection{Small amplitude excitations within Bogoliubov approximation}
\label{Bogol}
The next step is to consider small perturbations around the gray soliton background. Small amplitude excitations of any wave length in a Bose condensed gas may be described within the Bogoliubov linearized approximation. We assume that the field $\psi$ can be written as
\begin{equation}
\label{linearize}
\psi(x,t)= \psi_0(x)+ \varepsilon \!\;  \psi_1(x,t) \; ,
\end{equation}
where $\varepsilon$ is a small perturbation parameter and the soliton background  $\psi_0$ and the excitation field $ \psi_1$ are assumed to be of the same order of magnitude. We do a perturbation expansion in $ \varepsilon$ and obtain the BdGE for $ \psi_1$ as the linearization of the NLSE around the gray soliton background:
\begin{equation}
\label{bdg}
\mathrm{i} \partial_t \psi_1=\! \left (\! -\frac{1}{2} \partial_x^2 + \mathrm{i} (\beta-v) \partial_x + 2 | \psi_0|^2 -\tilde{\mu} \! \right)\! \psi_1 + \psi_0^2 \psi_1^{ \ast} \, ,
\end{equation}
in the frame co-moving with the soliton. A solution of the BdGE can be obtained in terms of a normal mode expansion with respect to complex variables $a_k(t),a_k(t)^ \ast$, labelled with mode index $k$. From the results in \cite{Walczak2011} one finds that 
\begin{equation}
\label{exc}
\psi_1(x,t) = e^{-\mathrm{i} v x} \int_{-\infty}^{\infty} \mathrm{d} k (u_k(x) a_k(t)+v_k^\ast (x) a_k^ \ast (t))
\end{equation}
is a solution to the BdGE \eqref{bdg}, where the time evolution of the complex variables is harmonic, i.e.,
\begin{equation}
a_k(t)= a_k e^{-\mathrm{i} \Omega_k t}, \quad a_k^\ast(t)= a_k^\ast e^{\mathrm{i} \Omega_k t} \; ,
\end{equation}
and the mode function $u_k(x),v_k(x)$ are given by
\begin{equation}
\label{uv}
\begin{split}
\begin{bmatrix} u_k \\ v_k \end{bmatrix} &=e^{\mp \mathrm{i} v x}e^{ \mathrm{i} k x} \frac{\mathrm{N}_k}{\Omega_k} \bigg[k \kappa^2 \operatorname{sech}^2\! \kappa x-2 \beta \Omega_k \\
& \qquad \qquad +(k^2 \pm 2 \Omega_k) \left(\frac{k}{2}+ \mathrm{i} \kappa \tanh \kappa x \right) \bigg]
\end{split}
\end{equation}
Here the upper (lower) signs of $\mp$ and $\pm$ apply for $u_k$ ($v_k$). The frequency $\Omega_k=-\beta k+\sqrt{k^4/4+ \mu k^2}$ is the Bogoliubov frequency for gray-soliton background and the normalization constant $ \mathrm{N}_k$ is chosen such that the mode functions are normalized to Dirac $\delta$-functions according to \cite{Fetter1972}. From the form of the exact solution one can see that excitations are not reflected by the soliton: it is absolutely transparent. The mode functions $u_k,v_k$ constitute the non-zero frequency part of a complete set of functions. For our considerations in this paper we will not need to introduce explicitly the modes with zero frequency, which require a slightly different treatment \cite{Walczak2011}. 

In what follows we will construct wave packets by specifying the amplitudes $a_k$ as a Gaussian envelope, but before doing so, we will examine the asymptotic behavior of the mode functions \eqref{uv} to the left and right of the soliton, since this will offer a basic insight into the non-trivial motion of perturbation wave packets in the soliton background. As will then become apparent in Sec.~\ref{creation}, this is in turn the origin of the soliton back-reaction displacement.

%%%%%%%%%%%%%%%%%%%%%%%%%%%%%%%%%%%%%%%%%%%%%%
\subsection{Asymptotic behavior of the normal mode functions}
\label{asymptotic}
To understand the essential feature of the asymptotic behavior of our $u_{k}$ and $v_{k}$, it is helpful to define the corresponding asymptotic limits of $\psi_{0}$ itself. Because this notion of asymptotic behavior away from the soliton will be basic to our paper we introduce the notation for any function $f(x)$
\begin{eqnarray}
	f_{>}(x)&\equiv& \lim_{e^{-2\kappa x}\to 0} f(x)\qquad\mathrm{for }\  x>0\nonumber\\
 	f_{<}(x)&\equiv& \lim_{e^{+2\kappa x}\to 0} f(x)\qquad\mathrm{for }\   x<0\nonumber\\
	\mathrm{and }\ f_{\gtrless}&\equiv& \lim_{e^{\mp2\kappa x}\to 0} f(x)
\end{eqnarray}
as a compact form to refer to either of the asymptotic cases alternatively. Referring to $\psi_{0}(x)$ as given by (\ref{sol}), we abbreviate $\psi_{0\gtrless}$ as simply $\psi_{\gtrless}$ and recognize that $\psi_{\gtrless} = e^{- \mathrm{i} v x}(\mathrm{i} \beta \pm \kappa)$. 

For $x>0$ the gray soliton solution \eqref{sol} is equal to $ \psi_>$ up to errors $ \mathcal{O}(e^{-2 \kappa |x|})$, whereas for $x<0$ it coincides with $ \psi_<$ up to $ \mathcal{O}(e^{-2 \kappa |x|})$. Both these wave functions $\psi_{>}$ and $\psi_{<}$ represent uniform, though not real, solutions to the NLSE \eqref{GP1}, with constant density. In particular both $\psi_{>}$ and $\psi_{<}$ have the same density $|\psi_{\gtrless}| = \sqrt{\kappa^{2}+\beta^{2}}\equiv\sqrt{\mu}\equiv c$, and the same phase gradient factor $e^{-\mathrm{i} v x}$; they differ only by a constant phase. It is easy to show that the finite frequency Bogoliubov solutions \eqref{uv} for the soliton background reduce, in the same $e^{\mp2\kappa x}\to0$ sense, to Bogoliubov solutions for the corresponding uniform backgrounds $ \psi_\gtrless$ \cite{Walczak2011}. The interesting point is precisely how the full solutions interpolate between these two asymptotic regions.

We will therefore compare our exact Bogoliubov solutions for the soliton background to the corresponding Bogoliubov solutions for a uniform, soliton-free background with the same modulus and phase gradient as $\psi_{\gtrless}$, namely $\bar\psi \equiv \sqrt{\kappa^{2}+\beta^{2}}e^{-\mathrm{i}vx}$. We introduce the functions $\bar{u}_{k}, \bar{v}_{k}$,
 \begin{equation}
 \label{bar}
 \bar{u}_{k} \pm \bar{v}_{k} \equiv \frac{1}{\sqrt{2 \pi}} \left(\frac{k^2/2}{\Omega_{k}+\beta k}\right)^{\pm \frac{1}{2}} \; ,
\end{equation}
which, when multiplied by phase factors $ e^{\mathrm{i} k x }$ and $e^{-\mathrm{i} v x}$, are exact Bogoliubov solutions for the uniform background solution $\bar\psi$ \cite{Pethick2008}.

Inserting the above definitions and applying some algebra, one finds that the Bogoliubov modes \eqref{uv} can be re-written as follows:
 \begin{equation}
 \label{asympt}
 \begin{bmatrix} u_k \\ v_k\end{bmatrix} 
 = e^{\frac{\mathrm{i}}{2}\theta_{k} \operatorname{sgn}(x)}\frac{e^{ \mathrm{i} k  x} \operatorname{sgn}(k)}{|\psi_\gtrless|} \begin{bmatrix} \bar{u}_{k} \psi_\gtrless 
 \\ \bar{v}_{k} \psi_\gtrless^* \end{bmatrix} +\mathcal{O}(e^{-2 \kappa |x|})\; ,
\end{equation}
where the phase $\theta_{k}$ is given by
 \begin{equation}
 \label{theta}
e^{\mathrm{i} \theta_k}  \equiv \frac{2 \mathrm{i} (k \kappa^2-\beta \Omega_k) +\kappa k^2}{2 \mathrm{i} (k \kappa^2-\beta \Omega_k)-\kappa k^2} \; .
\end{equation}
What we have thereby shown is that exact solutions in the soliton background reduce to uniform solutions on either side of the soliton, but with a particular $k$-dependent phase shift between the two asymptotic solutions, across the intervening soliton. We observe, for instance, the limiting cases
\begin{eqnarray}
	\lim_{k\to\pm\infty}e^{ \mathrm{i} \theta_{k}}&=&  \frac{\psi_<}{\psi_>} \nonumber\\
	\lim_{k\to 0}e^{ \mathrm{i} \theta_{k}}&=& 1\;.
\end{eqnarray}
This implies that for very large $k$ (for which $\bar{v}_{k}\to 0$) there is no order $\varepsilon$ phase shift across the soliton in $\psi = \psi_{0}+\varepsilon \psi_1$, while for small $k$ this total phase shift is simply given by the background phase shift that is already present in $\psi_{0}$. As a function of $k$, we can say that the excitation phase shift $\theta_{k}$ thus interpolates between two different senses of being trivial. When we consider the propagation of wave packets composed from our exact Bogoliubov excitations, however, we will see that $\theta_{k}$ has non-trivial effects for all $k$.

%%%%%%%%%%%%%%%%%%%%%%%%%%%%%%%%%%%%%%%%%%%%%%
%
%%%%%%%%%%%%%%%%%%%%%%%%%%%%%%%%%%%%%%%%%%%%%%
\section{Displacement of excitation pulses}
\label{creation}
\subsection{Bogoliubov wave packets}
From the solutions \eqref{exc} we will construct a class of Gaussian wave packet excitations.  In the limit of short wave lengths these will reduce to ordinary Schr\"odinger wave packets of free particles with velocity large compared to the speed of zero sound in the condensate background through which they move. In the limit of long wavelengths they will rather represent pulses of zero sound, broad enough to have well defined wavelength. For intermediate wavelength these packets are simply localized travelling excitations of an intermediate nature.

We define our wave packets by choosing
\begin{equation}
\label{amplitudes}
a_{k+\Delta k}= \frac{1}{\sqrt{2 \pi}}\, e^{-\frac{\lambda^2}{2} \Delta k^2}
\end{equation}
where the free parameters $k$ and $ \lambda$ define the mean inverse wavelength and total spatial extent of the wave packet, respectively.  For the following we define the dimensionless wavenumber difference $\zeta$ such that 
\begin{equation}
\Delta k=\zeta/\lambda
\end{equation}
and assume that $\lambda$ is large enough that \eqref{amplitudes} is a Gaussian distribution sharply centered around $\Delta k=0$. 

We explicitly compute the resulting excitation \eqref{exc} for $\kappa |x| \gg 1$, i.e., in the domain outside the soliton, by expanding $u_{k+\zeta/\lambda}$ and $v_{k+\zeta/\lambda}^*$, as they appear in the integrand of \eqref{exc}, in powers of $1/\lambda$.
Inserting \eqref{amplitudes} into \eqref{exc} and expanding the integrand to first order in $\lambda^{-1}$ we find by integration
\begin{align}
\label{exc11}
\psi_{1}^{\gtrless}\!  & \equiv \frac{\psi_\gtrless}{| \psi_\gtrless|} \frac{ e^{-\frac{1}{2 \lambda^2}z_\pm^2}}{ \lambda} \bigg[ e^{ \mathrm{i}(k x- \Omega_{k} t \pm \frac{\theta_k}{2})} \! \bigg(\!\bar{u}_k+ \mathrm{i} \frac{z_\pm}{\lambda^2}\bar{u}'_{k} \! \bigg)\\
\nonumber & \qquad \qquad \qquad+e^{- \mathrm{i}(k x- \Omega_{k} t \pm \frac{\theta_k}{2})} \! \bigg(\!\bar{v}_k- \mathrm{i} \frac{z_\pm}{\lambda^2} \bar{v}'_{k}\! \bigg) \! \bigg] \; , \\
\label{z} z_\pm &\equiv x-\left(\nu_k t \pm \frac{\Delta_k}{2} \right) \; ,
\end{align}
where we have introduced the $k$-dependent group velocity 
\begin{equation}
\label{group}
\nu_k \equiv \frac{d \Omega_{k}}{dk}=\operatorname{sgn}(k) \frac{k^2+2 c^2}{\sqrt{k^2+4c^2}}-\beta
\end{equation}
and position shift
\begin{equation}
\label{Deltak}
	\Delta_k \equiv -\frac{d \theta_{k}}{dk} = \frac{ \kappa k^2}{\Omega_k ( \Omega_k + \beta k)} 
\end{equation}
such that the envelope of the excitation pulse in \eqref{exc11} depends on the argument $z_{\pm}(x,t)$ as defined. In \eqref{exc11} and in the following  primes stand for derivatives with respect to the wavenumber $k$ unless otherwise stated.

We have hereby constructed a wave packet with well-defined wavenumber $k$ and Gaussian envelope of breadth $\lambda$, moving at group velocity $ \nu_k$. Since $|\beta|<c$, for $k>0$ ($k<0$) the group velocity $ \nu_k$ is positive (negative) and the envelope of the wave packet moves in positive (negative) $x$-direction. In the $|k| \gg1$ regime $\bar{v}_{k}\to 0$ and the excitation \eqref{exc11} reduces to a Sch\"odinger wave packet.

Eqs.~\eqref{exc11} and \eqref{z} imply that the wave packet has the relative spatial displacement $\pm\Delta_{k}/2$ between the two asymptotic situations at $x \gtrless 0$. A wave packet, which is initially located in the negative $x$-domain and has $k>0$, travels in positive $x$-direction and eventually passes the soliton; afterwards it will be seen to have been pushed forward by an extra distance $ \Delta_k$ in comparison to where it would have been if it had continued moving at constant group velocity, as it would have in the absence of the soliton. The wave packet displacement $\Delta_{k}$ is a monotonically decreasing function of $|k|$, and in the limiting cases obeys
\begin{align}
\label{dis0}\Delta_0 &= \frac{ \kappa}{c (c- \operatorname{sgn}(k) \beta)} \, ,\\
\label{diskk} \lim_{ k\to \pm \infty} \Delta_k &= \frac{4 \kappa}{k^2} \,.
\end{align}
The $k\to0$ limit in \eqref{dis0} can be understood in terms of the multiple scale theory of solitons interacting with general long-wavelength backgrounds \cite{Busch2000}, though the analysis from that perspective is somewhat involved, and may indeed perhaps be most easily understood by reversing our present viewpoint, and considering the advancement of the long wavelength packet as back-reaction from the soliton's motion. The high $k$ limit, however, can be easily understood semiclassically.

When the wavenumber $k$ is much greater than the inverse healing length of the condensate (set to 1 in our units), one can neglect $v_k$ and the BdGE reduces to a time-independent Schr\"odinger equation for $u_k$ \cite{Pethick2008}:
\begin{equation}
\label{semiclassical}
\left (-\frac{1}{2} \partial_x^2 + \mathrm{i} \beta \partial_x + V(x) \right) u_k(x) e^{\mathrm{i} v x} = E u_k(x) e^{\mathrm{i} v x} \,.
\end{equation}
In this equation $E$ is the energy eigenvalue and $V(x)= -2 \kappa^2 \operatorname{sech}^2 \kappa x$ is the mean field potential $2|\psi_{0}|^{2}$ exerted by the soliton-carrying condensate on short wavelength particles. In this high $k$ limit we can solve \eqref{semiclassical} in the WKBJ semiclassical approximation and find
\begin{equation}
\label{solwkb}
u_k e^{\mathrm{i} v x} \propto \exp \! \left[\mathrm{i} k \! \left( x + \frac{2 \kappa^2 }{k^2} \! \int_{-\infty}^x \! \! \mathrm{d} y \operatorname{sech}^2 \! \kappa y \right) \right]
\end{equation}
while the energy eigenvalue is $ E_k=k^2/2 -k \beta$. The first summand in the exponent of \eqref{solwkb} is equal to the result for constant background solution and the second summand reproduces the phase shift \eqref{diskk} for large $k$ in the limit $x \rightarrow \infty$. This result \eqref{solwkb} has the obvious classical interpretation that a particle accelerates and then decelerates as it passes through the conservative potential well $V(x)$, and thus emerges with its initial speed unchanged, but having taken less time to cross the well than it would have at constant speed.

%%%%%%%%%%%%%%%%%%%%%%%%%%%%%%%%%%%%%%%%%%%%%%
\subsection{Numerical propagation of wave packets}
\label{numwave}
For comparison with the above analytical perturbation theory, we have also investigated the motion of the wave packets by numerically solving the NLSE \eqref{GP1}, in cases both with and without a soliton present, and for various initial wave packet conditions. These results show that our analytical formula (\ref{Deltak}) for the soliton-induced wave packet advance $\Delta_{k}$ to leading order in perturbation amplitude $\varepsilon$ and wave packet momentum width $\lambda^{-1}$ is accurate.

\begin{figure}[t]
\includegraphics[width=8.6cm, angle=0]{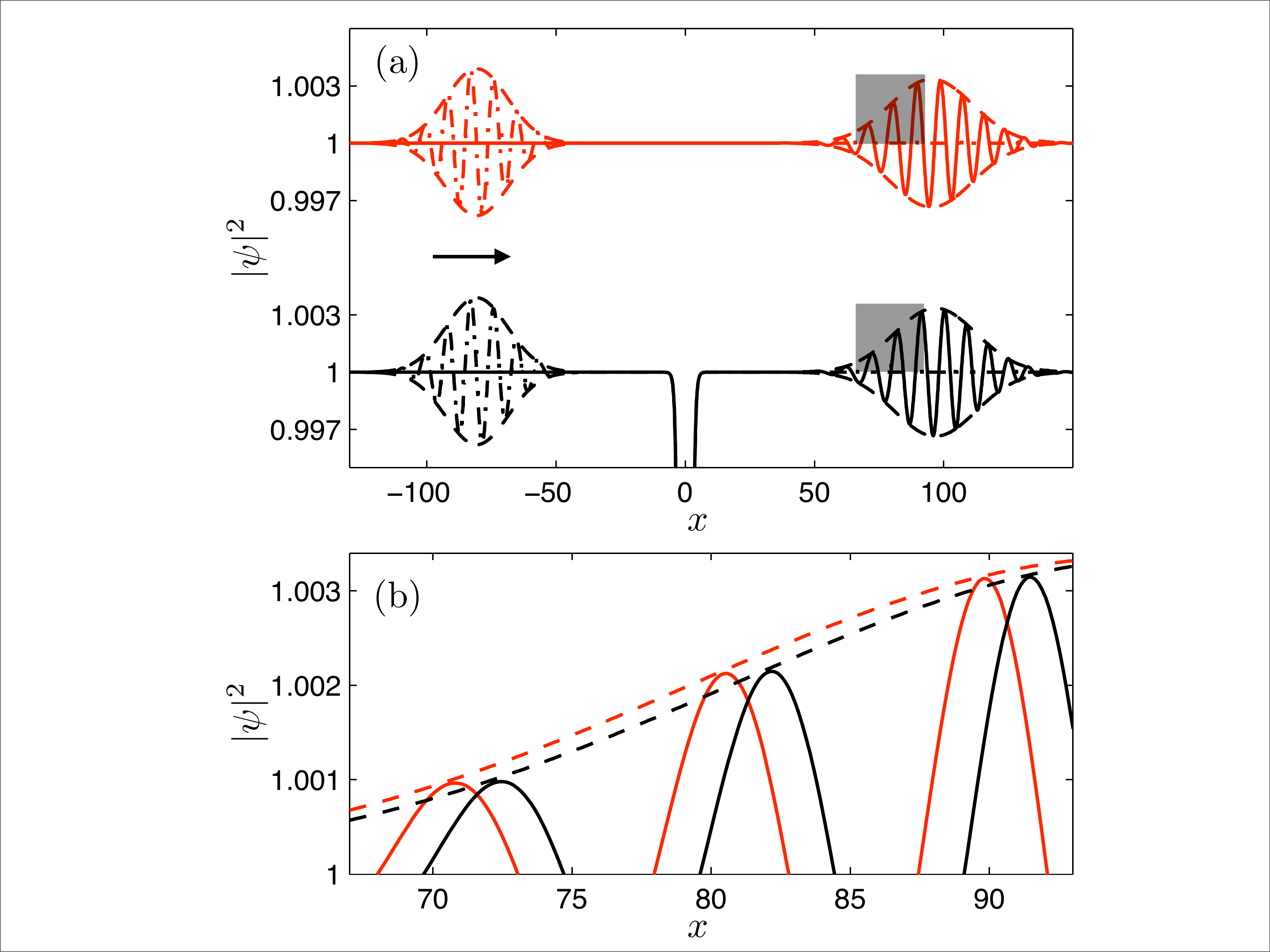}
\caption{(Color online) Wave packet displacement in a soliton background. In (a) the initial density of an excitation wave packet (dash-dotted line) characterized by $k=0.7$, $\lambda=12$ and $\varepsilon=0.02$ and center at $x=-80$ in a constant background (red lines) and in a soliton background (black lines) with the same asymptotic density and gas velocity is shown; the arrow indicates the propagation direction of the wave excitation. The soliton is located at $x=0$ and its other solution parameters are $\beta=v=0.5$ and $\sqrt{\mu}=1$. The solid lines represent the density after the same propagation time for both backgrounds. In dashed lines the analytical envelopes including higher order corrections in $\lambda^{-1}$ before and after the propagation are depicted. Fig.~(b) is a magnification of the gray square area in (a) and shows the final density of the constant background (red) and soliton background (black) with the corresponding analytical envelopes in dashed lines.}
\label{figdis}
\end{figure}

In Fig.~\ref{figdis} we show a typical example for the propagation of the wave packet excitation. We compare the result for constant background $\bar{\psi}$ to a soliton background ($ \beta=v=0.5$ and $\sqrt{\mu} =1$) with equal asymptotic density and gas velocity. The perturbation amplitude parameter is $\varepsilon=0.02 \ll1$. At initial time the center of a wave packet with mean wavenumber $k=0.7$ and spatial breadth $\lambda=12$ is placed at $x=-80$ in both backgrounds, such that at initial time the envelopes $|\psi_0+\varepsilon \psi_1|^2$ are identical. Under time evolution the wave packets move in the positive $x$-direction. The numerical integration was extended to the same final time for both backgrounds, at which point the excitation had in both cases passed well beyond the position where the soliton is in the case where it is present. We compare these numerically exact propagations of the wave packets to the analytically predicted motion of its Gaussian envelope, as given by the wave packet group velocity and the soliton-induced shift $\Delta_{k}$. Our analytical envelope motion has also included order $\lambda^{-2}$ corrections to the group velocity, as described in Appendix \ref{secondlambda}, since our total evolution time is long enough that this small velocity correction produces a noticeable shift in the packet position at late times. This group velocity correction is the same whether or not the soliton is present, and all other order $\lambda^{-2}$ corrections are too small to be seen in our plots.

%%%%%%%%%%%%%%%%%%%%%%%%%%%%%%
%FIG: wave packet shift
%%%%%%%%%%%%%%%%%%%%%%%%%%%%%%
%%%%%%%%%%%%%%%%%%%%%%%%%%%%%%
%FIG: wave packet shift
%%%%%%%%%%%%%%%%%%%%%%%%%%%%%%

We have also numerically confirmed the wave packet displacement $\Delta_{k}$ for many values of $k$ and soliton speed $\beta$ other than those represented in Fig.~\ref{figdis}. Since for short wavelength packets, according to \eqref{diskk},  the packet displacement soon becomes even shorter than the wavelength, it is clear that the soliton-induced displacement of short-wavelength pulses is an all but negligible effect. A soliton may be passed, however, by arbitrarily many excitation packets. The cumulative back-reaction effect on the soliton can thus become arbitrarily large. To this back-reaction effect we now turn.

%%%%%%%%%%%%%%%%%%%%%%%%%%%%%%%%%%%%%%%%%%%%%%
%
%%%%%%%%%%%%%%%%%%%%%%%%%%%%%%%%%%%%%%%%%%%%%%
\section{Soliton displacement as a second order process}
\label{secondorderprocess}
At first order in $ \varepsilon$ we have shown that the wave packet is pushed forward by the soliton, while the soliton itself is asymptotically unaffected. The reason for qualifying the soliton's invariance as `asymptotic' is that for perturbations with wavelength longer than the soliton width, the terms in $u_{k}(x)$ and $v_{k}(x)$ that are proportional to $\operatorname{sech}^2\!\kappa x$ have exactly the effect of translating the soliton (since they are proportional the spatial derivative of $\psi_{0}(x)$). The terms in $u_{k}(x)$ and $v_{k}(x)$ proportional to $\tanh\kappa x$ likewise have effects on $\psi=\psi_{0}+\varepsilon\psi_{1}$ that are indistinguishable, for small $k$, from perturbations of $\beta$ in $\psi_{0}$. By performing our time-dependent wave packets' Gaussian integrals, however, one can directly show that all these soliton perturbations vanish except when the wave packet envelope is near to the soliton. There is no lasting effect on the soliton, at order $\varepsilon$, from Gaussian wave packets of the form we discuss.

The soliton cannot possibly remain exactly unaffected by a wave packet, however, even asymptotically. A wave packet carries a finite total density perturbation, and thus makes a finite contribution to the motion of the system's center of mass. Momentum conservation and continuity (particle number conservation) together imply that the system's center of mass must travel at an exactly constant speed. The forward jump by the excitation pulse as it passes the soliton contributes a brief interval of acceleration to the center of mass. It is therefore inconsistent with the exact conservation laws for the soliton to act on the wave packet in this way, without any corresponding back-reaction.

This apparent center of mass jump is not a contradiction to our exact order $\varepsilon$ results, because the wave packet only contributes to the center of mass at order $\varepsilon^2$. So what we have deduced from continuity and momentum conservation is a constraint on the order $\varepsilon^2$ post-Bogoliubov corrections: they must cancel the jump of the center of mass that is contributed by the wave packet jump $\Delta_{k}$. As we will see, the back-reaction displacement of the soliton is one of these demonstrably necessary $\varepsilon^2$-corrections. But to obtain the post-Bogoliubov soliton back-reaction correctly, we must consider the corrections at post-Bogoliubov order generally.

%%%%%%%%%%%%%%%%%%%%%%%%%%%%%%%%%%%%%%%%%%%%%%
\subsection{Wave function at order $\varepsilon^2$}
\label{2ndordersolution}
We expand the NLSE \eqref{GP1} up to second order in $\varepsilon$ with
\begin{equation}
\label{secondo}
\psi= \psi_0(x)+ \varepsilon \!\; \psi_1(x,t)+ \varepsilon^2 \psi_2(x,t)
\end{equation}
where $\psi_2$ is assumed to be of the same order of magnitude as $ \psi_0$ and $ \psi_1$. At order $ \varepsilon^2$ the NLSE is then given by
\begin{multline}
\label{e2}
\mathrm{i} \partial_t \psi_2= \! \left(\! -\frac{1}{2} \partial_x^2 + \mathrm{i} (\beta-v) \partial_x +2 | \psi_0|^2 - \tilde{ \mu}\!  \right) \! \psi_2+ \psi_0^2 \psi_2 ^ \ast\\ + 2|\psi_1|^2 \psi_0+ \psi_0^ \ast \psi_1^2  \, .
\end{multline}
It is of the same form as the BdGE \eqref{bdg} except for the driving term on the right-hand side of Eq.~\eqref{e2}, provided by the soliton background and the first order contribution $\psi_1$. The solution to the homogeneous part of (\ref{e2}) is therefore again of the form \eqref{exc}. To find a particular solution of \eqref{e2} we expand the second order piece $ \psi_2$ of the condensate wave function in terms of a complete set of functions:
\begin{equation}
\label{psi2}
\begin{split}
\psi_2(x,t) &= \int \! \mathrm{d} l  \{u_l(x) b_l(t)+v_l^\ast (x) b_l^ \ast (t)\} \\
& \qquad + q_z(t) R_z(x)+ \mathrm{i} p_z(t) S_z(x)
\end{split}
\end{equation}
where $b_l(t)$ and $b_l^\ast(t)$ are complex functions of time and $q_z(t)$ and $p_z(t)$ are real functions of time. The continuum mode functions alone  $\{u_l,v_l\}$ do not constitute a complete set of functions; we must also include the discrete zero mode with negative mass $m_z=-4 \kappa$ \cite{Negretti2008a}, associated with the spatial translation of the soliton, to achieve completeness \cite{Walczak2011}. The discretely normalized mode functions in \eqref{psi2} are 
\begin{equation}
\label{rzsz}
\begin{split}
\begin{bmatrix} R_z\\S_z \end{bmatrix} &=\begin{bmatrix} (\mathrm{i} v+ \partial_x) \psi_0 \\ \mathrm{i} \partial_\beta \psi_0/m_z \end{bmatrix} \\
&= e^{-\mathrm{i} v x} \begin{bmatrix} \kappa^2 \operatorname{sech}^2 \kappa x \\ (\kappa +\mathrm{i} \beta(\tanh \kappa x+ \kappa x \operatorname{sech}^2 \kappa x))/(4\kappa^2) \end{bmatrix} \,.
\end{split}
\end{equation}
The representation of the discrete zero mode with real variables $q_{z}$, $p_{z}$, rather than the usual complex co-efficients related to quantum mechanical creation and destruction operators, is necessary because harmonic oscillator raising and lowering operators (and their classical counterparts) are singular for the free-particle limit of a harmonic oscillator.

Since we have explicit analytic solutions to the homogeneous equation (\ref{bdg}), it is straightforward to construct the Green's function to solve (\ref{e2}). In fact we need simply insert (\ref{psi2}) into (\ref{e2}) and use the fact that $u_{l}$ and $v_{l}$ are orthonormal solutions to (\ref{bdg}) to obtain the readily integrable first order differential equation for the $b_l(t)$:
\begin{equation}
\label{bkdif}
\begin{split}
&(\mathrm{i} \partial_t- \Omega_l) b_l(t)\\
=& \! \int \! \! \mathrm{d} x [ u_l^\ast (2 \psi_0 |\psi_1|^2+ \psi_0^ \ast \psi_1^2)+v_l^\ast (2 \psi_0^\ast |\psi_1|^2+ \psi_0 \psi_1^{\ast 2})] \; ,
\end{split}
\end{equation}
A particular solution is always offered by 
\begin{widetext}
\begin{equation}
\label{solbl}
\begin{split} 
b_{l}(t) &= \int\frac{\mathrm{d} l' \mathrm{d} l''}{2\pi \mathrm{i}}e^{-\frac{\lambda^{2}}{2}[(l'-l)^{2}+(l''-l)^{2}]}\\
&\qquad \times\Big\{\frac{e^{-\mathrm{i} (\Omega_{l'}+\Omega_{l''})t}}{\Omega_{l}-\Omega_{l'}-\Omega_{l''}}\int\!\mathrm{d} x\, [u_{l}^{*}u_{l'}(2\psi_{0}v_{l''}+\psi_{0}^{*}u_{l''})+v_{l}^{*}v_{l'}(2\psi_{0}^{*}u_{l''}+\psi_{0}v_{l''})]  \\
&\qquad \quad +\frac{e^{ \mathrm{i} (\Omega_{l'}+\Omega_{l''})t}}{\Omega_{l}+\Omega_{l'}+\Omega_{l''}}\int\! \mathrm{d} x\,[u_{l}^{*}v_{l'} (2\psi_{0}u_{l''}^{*}+\psi_{0}^{*}v_{l''}^{*})+v_{l}^{*}u_{l'}^{*}(2\psi_{0}^{*}v_{l''}^{*}+\psi_{0}u_{l''}^{*})] \\
& \quad\qquad + \frac{e^{-\mathrm{i}(\Omega_{l'}-\Omega_{l''})t}}{\Omega_{l}-\Omega_{l'}+\Omega_{l''}}\int\! \mathrm{d} x\,[u_{l}^{*}u_{l'} (2\psi_{0}u_{l''}^{*}+\psi_{0}^{*}v_{l''}^{*})+u_{l}^{*}v_{l''}^{*}(2\psi_{0}v_{l'}+\psi_{0}^{*}u_{l'}) \\
&\quad \qquad\qquad \qquad\qquad\qquad\qquad+v_{l}^{*}u_{l''}^{*}(2\psi_{0}^{*}u_{l'}+\psi_{0}v_{l'})+v_{l}^{*}v_{l'} (2\psi_{0}^{*}v_{l''}^{*}+\psi_{0}u_{l''}^{*}) ]\Big\} \;.
\end{split}
\end{equation}
\end{widetext}
(Here the prime on the wavenumber $l$ does not indicate any differentiation, but simply distinguishes $l'$ and $l''$ from $l$ as an integration variable.) Any solution $b_{l}(t)\propto e^{-\mathrm{i}\Omega_{l}t}$ to the homogeneous equation can also be added to this particular solution, to satisfy initial conditions. Similar integrals provide exact particular solutions for $q_{z}(t)$ and $p_{z}(t)$.

These integrals \eqref{solbl} can be evaluated exactly if need be, since all the $x$-integrals can be performed for any $t$; but by considering the $e^{\mathrm{i} l x}$ prefactors in $u_l$ and $v_l$, we can see that for early and late times, respectively before and after the wave packet has passed the soliton, the integrals over $l'$ and $l''$ evaluate to wave packets in $x$ that are Gaussianly concentrated well before or after the soliton. (This is obvious, inasmuch as we know that $\psi_{1}$ is a wave packet that propagates through the soliton without reflection.) For $t$ either before or after the packet passes the soliton, therefore, we can evaluate \eqref{solbl} to essentially perfect accuracy by replacing the $\psi_0(x)$, $u_l(x)$ and $v_l(x)$ functions in its integrals over $x$ with their asymptotic forms for $\kappa|x|\gg 1$. This makes the $x$-integrals in \eqref{solbl} quite straightforward, for early and late $t$, allowing all co-efficients $b_{l}(t)$, $q_{z}(t)$ and $p_{z}(t)$ to be computed quite simply, except during the brief time while the wave packet overlaps with the soliton. (The behavior of these coefficients during the overlap interval can also be determined, but the complicated expressions in this case do not seem to us to add any conceptual understanding, beyond the qualitative fact that all the co-efficients change during this time.)

We can then obtain explicit expressions for the $l$ integrals in $\psi_{2}$, for all $t$ either before or after the wave packet has passed the soliton. In general these are again rather complicated, since $u_{l}$ and $v_{l}$ are non-trivial; but it is again apparent from the integrand's form (for $t$ before or after the soliton-packet overlap) that $\psi_{2}$ does not include any new wave packets, either reflected or otherwise emitted. Rather, $\psi_{2}$ includes just two kinds of contributions. 

Firstly, the single propagating wave packet is dressed: it acquires additional components, at order $\varepsilon^{2}$, that are either higher or lower harmonics of the order-$\varepsilon$ waveform $\sim e^{\pm \mathrm{i}kx}$. The higher harmonic component is a co-traveling wave packet with mean wave number $2k$, while the lower harmonic component is a smooth pulse whose only spatial scale is the packet breadth $\lambda$. Both of these harmonic packets travel with the primary packet, sharing its group velocity and dispersion; they do not separate from it to move at the speeds at which isolated disturbances of their respective forms would propagate. Away from the soliton, this post-Bogoliubov dressing of the $\psi_{1}$ wave packet reduces to exactly the same form that one finds if the calculation is repeated with a soliton-free background $\psi_{0}$, except that with the soliton the wave packet is displaced by the same $\Delta_{k}$ as derived above:
\begin{multline}
\label{exc2}
\psi_{2}^ \gtrless = \frac{ \psi_\gtrless }{((\nu+\beta)^2-c^2)}  \left[ \frac{e^{-z_\pm^2/2 \lambda^2}}{\lambda^2}( \eta_k+(\nu+\beta) \zeta_k) \right. \\
 + \mathrm{i} \frac{\sqrt{\pi}}{ \lambda} \operatorname{erf} \left(\frac{z_\pm}{\lambda}\right) ((\nu+\beta) \eta_k+c^2 \zeta_k) \Bigg] \\
 +\psi_2^{\mathrm{fast}} + \mathcal{O}(\lambda^{-3})
\end{multline}
where we have defined the quantities
\begin{equation}
\label{eta}
\begin{split}
\eta_k &\equiv (| \bar{u}_k |^2+ |\bar{v}_k|^2+\bar{u}_k \bar{v}_k) \, ,\\
\zeta_k &\equiv (\bar{v}_k \bar{u}'_k-\bar{u}_k \bar{v}'_k) \, ,
\end{split}
\end{equation}
and used the definition $z_\pm = x-\nu_{k}t\mp\Delta_{k}/2$ from \eqref{z}. In \eqref{exc2} we have explicitly displayed only the smooth dressing of the lower harmonic component; $\psi_2^{\mathrm{fast}}$ denotes the higher harmonic terms proportional to the fast spatial oscillations $e^{\pm2 \mathrm{i} k x}$, whose explicit form will not be needed, for reasons that will be clear in our next Section. The lower harmonic dressing includes a phase perturbation with amplitude of order $\lambda^{-1}$, whose spatial profile is the integral of the wave packet's Gaussian envelope. This phase perturbation thus has a profile proportional to the error function, which has the property that $\operatorname{erf}(\infty)=\operatorname{erf}(-\infty)+1$. This means that a broad Bogoliubov--de Gennes wave packet at any finite $k$ permanently shifts the phase of $\psi$ behind it as it passes, by a constant of order $\varepsilon^2/\lambda$. It is perhaps surprising that a localized wave packet, even with high $k$, has in this sense a long range effect. But this is simply due to the fact that the sub-harmonic post-Bogoliubov components vary on the length scale $\lambda$ of the wave packet envelope, and so for $\lambda$ much longer than the healing length, these spatially smooth subharmonic corrections are in the hydrodynamic regime of the nonlinear Schr\"odinger dynamics, in which density pulses are necessarily accompanied by phase steps. This subharmonic dressing effect is present, unchanged, in the absence of solitons.

The second contribution in $\psi_{2}$ is one that persists in the vicinity of the soliton after the wave packet has passed it: the soliton is slightly displaced. It is otherwise exactly unchanged from its initial state, once the wave packet has passed it. This displacement is the back-reaction that we have been seeking, and in total it comes from two sources. Firstly there is a non-vanishing contribution from the continuum of $b_{l}$ modes, which settles to a constant value even after the wave packet has passed, because the term in the integrand of (\ref{psi2}) proportional to $\operatorname{sech}^{2}\! \kappa x$ has finite weight at small $l$. And secondly there is a contribution from the discrete soliton zero mode. The zero mode momentum coefficient $p_{z}(t)$ returns to its initial value after the wave packet has passed, and this initial value may be set to zero without loss of generality since an initial $p_{z}$ can be absorbed into the $\beta$ of $\psi_{0}$. But the soliton displacement amplitude $q_{z}(t)$ changes from zero to a non-zero value as the wave packet passes, and remains at this constant value thereafter. 

Both of these contributions to the soliton displacement back-reaction can in principle be computed directly, with $q_{z}$ and $p_{z}$ obtained by solving their own equations of motion,
\begin{eqnarray}\label{}
	\dot{p}_{z} &=& - 2\,\mathrm{Re} \!\int\! \mathrm{d}x\,R_{z}^{*}[2|\psi_{1}|^{2}\psi_{0}+\psi_{1}^{2}\psi_{0}^{*}]\nonumber\\
\dot{q}_{z} &=& \frac{p_{z}}{4\kappa} + 2 \, \mathrm{Im}\!\int\!\mathrm{d}x\,S_{z}^{*}[2|\psi_{1}|^{2}\psi_{0}+\psi_{1}^{2}\psi_{0}^{*}]
\end{eqnarray}
which together form the $z$-mode analog of (\ref{bkdif}), namely the projection of (\ref{e2}) onto the zero mode subspace of $\psi_{2}(x,t)$ that is spanned by $R_{z}(x)$ and $S_{z}(x)$. The modest difficulty of solving these coupled equations directly may be avoided, however, by using an indirect way of computing the soliton back-reaction, that is nonetheless just as exact as the direct approach, because it is based on exact conservation laws of the NLSE.

%%%%%%%%%%%%%%%%%%%%%%%%%%%%%%%%%%%%%%%%%%%%%%
\subsection{Soliton displacement from conservation laws}
\label{soldisplace}
Having explicitly solved the NLSE in the domain $\kappa |x| \gg 1$ to second order in $\varepsilon^2$, we can now determine the back-reaction on the soliton from exact conservation of momentum and particle number.  Since the NLSE conserves both of these quantities exactly, its expansion in powers of $\varepsilon$ necessarily conserves them at each individual order in $\varepsilon$. Both the wave packet dressing and soliton back-reaction contributions in $\psi_{2}$ contribute to total momentum and particle number at order $\varepsilon^{2}$, and so from our exact expressions for $\psi_{1}$ and the dressing part of $\psi_{2}$, we can use the conservation laws to infer the back-reaction contributions.

To this end we define the center of mass $Q$ and total linear momentum $P$ of the system
\begin{align}
\label{Q} Q &\equiv \int \! \mathrm{d} x \, x | \psi|^2\;, \\
\label{P} P &\equiv \frac{\mathrm{i}}{2} \int \! \mathrm{d} x \left( \psi \partial_{x} \psi^\ast - \psi^\ast \partial_x \psi \right) \;,
\end{align}
in the usual way. Using the NLSE \eqref{GP1} (in particular its imaginary part, corresponding to particle continuity) and integration by parts, the exact equation of motion for the center of mass can be established as
 \begin{equation}
 \label{motioncenter}
 \dot{Q}= P+(\beta-v) \int \mathrm{d} x |\psi|^2 \;.
 \end{equation}

It also follows readily from the NLSE that the total linear momentum of the system $P$ is exactly constant, as is the second term on the right hand side of \eqref{motioncenter}, which in the quantum gas context is associated with the total number of particles
\begin{equation}
\label{N} N \equiv \int \! \mathrm{d} x \, | \psi|^2\;.
\end{equation}
Consequently, the center of mass $Q$ moves at exactly the same speed at all times. We can easily compute this speed to order $\varepsilon^{2}$ during the evolution of our soliton and wave packet, by using the simple asymptotic forms for $\psi_{1}$ and $\psi_{2}$ when the excitation pulse is well localized outside the soliton domain. We can simplify the expressions without loss of generality by choosing the reference frame in which the soliton is at rest ($\beta=v$), and thereby obtain simply
\begin{align}
 \label{pexp}
 \dot{Q}&= \nu_k N+\mathcal{O} (\varepsilon^3) \;,
\end{align}
where we define the two contributions $N_1$ and $N_2$ to the total number of atoms $N= N_1 +N_2+\mathcal{O} (\varepsilon^3)$ by
\begin{align}
 \label{n1}
N_1&\equiv \varepsilon^2 \! \int \! \! dx \, | \psi_1^\gtrless|^2= \varepsilon^2 \frac{\sqrt{2 \pi}}{\lambda} ( \bar{u}_k^2+\bar{v}_k^2)\\
 \label{n2}N_2 &\equiv \varepsilon^2 \! \int \! \! dx \, (\psi_0 \psi_2^{\gtrless\ast}+ \psi_0^ \ast \psi_2^\gtrless)\\
\nonumber&= \varepsilon^2 \frac{\sqrt{2 \pi}}{\lambda}\frac{\bar{u}_k^2+ \bar{v}_k^2+\bar{u}_k \bar{v}_k+(\nu_k+\beta) (\bar{v}_k \bar{u}'_k-\bar{u}_k \bar{v}'_k)}{((\nu_k+\beta)^2-c^2)/(2 c^2)} \;.
\end{align}
In this case the velocity at which the center of mass propagates is determined by the group velocity $\nu_k$ of the excitation pulse and the integrated density of the excitation, \textit{i.e.}, the total number $N$ of excited atoms it contains, including both Bogoliubov and (first) post-Bogoliubov contributions. (We can now see why the precise form $\psi_{2}^{\mathrm{fast}}$ of the higher harmonic dressing components  $~ e^{ \pm \mathrm{i} 2 k x}$ in \eqref{exc2} is not needed: it makes no contribution to $N$ at order $\varepsilon^{2}$.)

Since the center of mass speed is exactly constant, for any times $t_\pm$ we must obviously have
 \begin{equation}
 \label{com}
 Q(t_+)-Q(t_-) =  (t_+-t_-) \nu_k N + \mathcal{O}(\varepsilon^3) \;.
 \end{equation}
But if we neglect the soliton back-reaction, and evaluate the left-hand side of \eqref{com} by inserting the asymptotic wave functions \eqref{exc11} and \eqref{exc2} in the definition \eqref{Q}, we find that it does not equal the right-hand side -- if $t_-$ denotes a time before the wave packet interacts with soliton, and $t_+$ a time at which the packet has passed it. Explicitly, we find
 \begin{equation}
 \label{com2}
 Q^{(>)}(t_+)-Q^{(<)}(t_-)= (t_+-t_-) \nu_k N+ \Delta_k N  \; ,
 \end{equation}
where with the superscripts $ >$ and $<$ we indicate that we have computed the center of mass using only the corresponding asymptotic wave functions away from the soliton, which include only the packet and its dressing, but not the soliton back-reaction. As we have seen, the soliton has pushed the excitation packet forward by an extra distance $\Delta_k$; and this has contributed an extra displacement of the center of mass at order $\varepsilon^2$ by $\Delta_k N$. 

This cannot be the only contribution at order $\varepsilon^{2}$ to the center of mass motion, for it is incompatible with the exact constant speed of the center of mass, as implied by the conservation laws. Hence we can infer that the soliton displacement contribution in $\psi_{2}$ must supply the compensating correction, in order to preserve the exact result \eqref{com}. Computing the left-hand side of \eqref{com} for the soliton background \eqref{sol}, where at $t_+$ we account for a shift $\Delta x$ in the soliton position by $x \to x-\Delta x$, we find that the soliton shift results in a center of mass displacement of
 \begin{equation}
 \label{com0}
Q^{(0)}(t_+)-Q^{(0)}(t_-) = -2 \kappa \Delta x \; .\\
 \end{equation}
Here the superscript $0$ indicates that we have inserted the gray soliton solution $ \psi_0$ in \eqref{Q}. The factor $-2 \kappa$ corresponds to the integrated linear density `missing' in comparison to the otherwise constant background; in this sense it is proportional to the (negative) mass of the soliton. The center of mass displacement from the dressed excitation wave packet and the soliton back-reaction must cancel each other due to exact conservation of total linear momentum, and therefore Eqs.~\eqref{com} to \eqref{com0} yield
 \begin{equation}
 \label{shift}
\Delta x = \Delta_k \frac{N}{2 \kappa} = \frac{k^2 (N_1+N_2)}{2 \Omega_k (\Omega_k+\beta k)} 
 \end{equation}
for the soliton displacement as back-reaction in post-Bogoliubov theory. Due to the negative mass $-2 \kappa $ of the soliton, the soliton back-reaction displacement is in the same direction as the wave packet displacement: the wave packet drags the soliton with it a short distance. Furthermore one can see from the definition of the particle numbers in \eqref{n1} and \eqref{n2} that the displacement is a second order effect in wave packet amplitude.
 
Equation \eqref{shift} is the main analytical result of this paper, and demonstrates the back-reaction on the soliton from wave excitations. It is determined by the finite shift of the excitations \eqref{Deltak} in the gray soliton background, and the ratio of atom number atom in the excitation pulse and the soliton. The limit behavior of $\Delta_k$, as displayed in \eqref{diskk}, implies that the back-reaction is maximal for $k=0$ and in the regime $k \gg 1$ drops quadratically in $k$ for fixed ratio $N/2 \kappa$. From the latter we conclude that short-wavelength particle-like excitations have a relatively small effect back-reaction effect on the soliton, whereas long-wavelength phase perturbations have the strongest interaction with the soliton.

%%%%%%%%%%%%%%%%%%%%%%%%%%%%%%%%%%%%%%%%%%%%%%
\subsection{Numerical analysis}
\label{NAdefinite}
The analytical formula \eqref{shift} for the soliton displacement is now compared to numerical solutions of the NLSE for periodic boundary conditions using a split step method \cite{Weideman1986}.  In this subsection we choose for the asymptotic density and speed of sound $ \mu=c^2=1$, and place the initial soliton at $x=0$. We emphasize that all results presented above for the infinite system are valid for the finite system to order $\mathcal{O}\left(e^{-2 \kappa L} \right)$; but a few complications of detail are introduced by embedding the soliton and wave packet in a finite geometry. 

For periodic systems there must in general be a background phase gradient $\psi_{0}\propto e^{-\mathrm{i}vx}$ in the presence of the soliton, and the allowed values for the background flow speed $v$ form a discrete set \cite{Walczak2011}. Within this constraint, however, we can easily fine tune the system size $2L$ to achieve $\beta=v$, so that the soliton is then (initially and finally) at rest in the laboratory frame, and the soliton displacement $\Delta x$ can easily be extracted from the density profile after time propagation. In the finite system, integrals over wavenumbers $k$ naturally become summations over the discrete solutions to 
\begin{equation}
\label{conditionk}
k\cot k L=2\left( \kappa-\frac{\beta \Omega_k}{\kappa k} \right ) \; ,
\end{equation}
according to \cite{Walczak2011}. By taking into account the freedom of an extra phase, which is allowed by the BdGE and the second order equation \eqref{e2}, we can place the center of the excitation pulse at initial time $t_i=0$ at any position $x_i$; we choose a starting point well separated from the soliton. This is equivalent to replacing $a_k \to a_k e^{-\mathrm{i} k x_i}$ in \eqref{amplitudes}. Furthermore we rescale our small perturbation parameter
\begin{equation}
\label{normeps}
\varepsilon \to \frac{\lambda}{\sqrt{2 \pi}( \bar{u}_k^2+\bar{v}_k^2)} \varepsilon
\end{equation}
such that $N_1$ is normalized to $ \varepsilon^2$. 

A less obvious complication in the periodic realization of our problem is that, once the order-$\varepsilon^{2}$ dressing is included, the total initial wave function that we have described in this paper heretofore is in general non-periodic, because of the phase step in \eqref{exc2}. To realize on the ring an equivalent encounter between wave packet and soliton, therefore, we add to our initial state a BdG solution of order $\varepsilon^2$ in the form of a hydrodynamic pulse \cite{Stringari1996} that cancels the total phase step around the ring, but then travels away from the soliton, and does not meet it or the main wave packet before the end of our numerical time evolution.
%%%%%%%%%%%%%%%%%%%%%%%%%%%%%%
%FIG: shift analysis short
%%%%%%%%%%%%%%%%%%%%%%%%%%%%%%
\begin{figure}[t]
\includegraphics[width=8.4cm, angle=0]{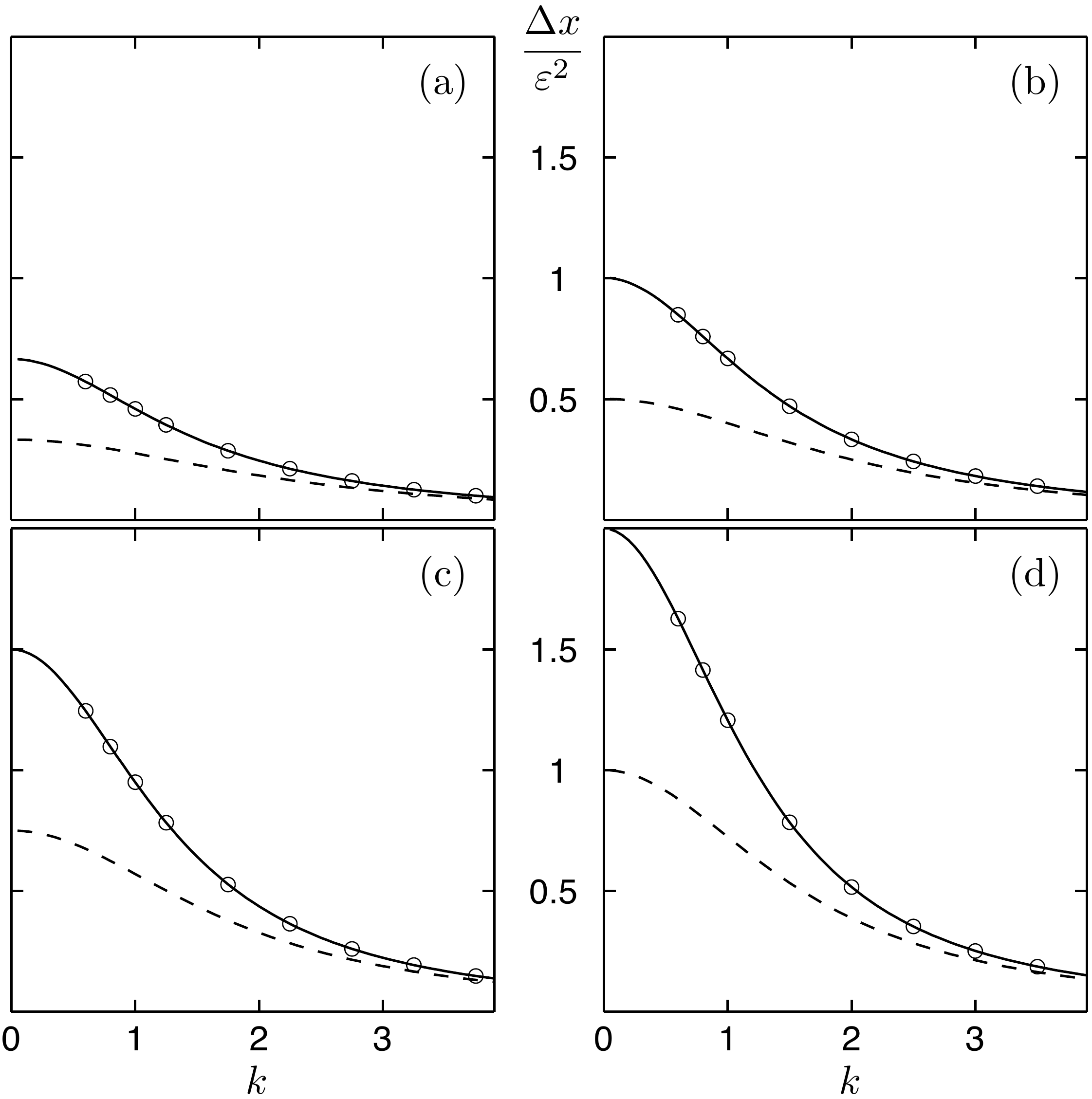}
\caption{Normalized soliton shift $\Delta x/\varepsilon^2$ as a function of the central wavenumber $k$ for different soliton speeds $\beta$. The data points represent averages of the normalized soliton shift over the pulse amplitudes $\varepsilon$, which have been varied from $0.03$ to $0.3$ in the simulations. The solid lines show the theoretical predictions with systematic inclusion of terms of order $\varepsilon^2$, while the dashed lines indicate the corresponding curves if second order contributions to the wave function $\psi$ are not taken into account. The soliton speed was given by (a) $\beta=-1/2$, (b) $\beta=-0.0058$, (c) $\beta=1/3$ and (d) $\beta=1/2$.}
\label{figshift}
\end{figure}
%%%%%%%%%%%%%%%%%%%%%%%%%%%%%%
%FIG: shift analysis short
%%%%%%%%%%%%%%%%%%%%%%%%%%%%%%
 
We have solved the time evolution of the wave function under the NLSE for different soliton speeds $\beta$ and for different sets of the excitation parameters $k$ and $\varepsilon$, keeping $\lambda=12$ fixed. The parameters obey the conditions $k>0$ and $k \lambda \gg1$, so that we have a wave packet of well defined wavelength moving in the positive direction. After the excitation has passed the soliton, the time evolution is stopped and the soliton position is extracted from the density profile of the final wave function \footnote{The nonlinear function $c_1^2+c_2^2 \tanh^2 \!c_2(x-c_3)$ is fitted to the density in the range $|x| \leq 15$. The fit parameter $c_3$ yields the soliton position.}. For each value of the wavenumber $k$ we have varied the amplitude parameter $\varepsilon$ between $0.03$ and $0.3$. The quantity $\Delta x/\varepsilon^2$ is expected to be independent of $\varepsilon^2$ from \eqref{n1}, \eqref{n2} and \eqref{shift}. Indeed, the relative deviation of this normalized soliton displacement from its average over the amplitudes never exceeds $0.2\%$ for any $k$. Thus, the numerical solutions clearly confirm the soliton displacement to be an effect quadratic in the amplitude. 

In Fig.~\ref{figshift} the numerical results for $\Delta x/\varepsilon^2$, averaged over all chosen amplitudes $\varepsilon$, are shown as circles, for different values of the wavenumber $k$ and soliton speed $\beta$. It is apparent that for fixed soliton speed the soliton displacement is a decreasing function of the wavenumber $k$, while for fixed wavenumber the displacement increases for increasing soliton speed. Furthermore Fig.~\ref{figshift} suggests that even excitations with central wavenumber $k$ larger than the inverse healing length, which is one in our dimensionless units, still have a noticeable impact on the soliton. The solid lines in Fig.~\ref{figshift} indicate the analytical behavior, including all contribution to the density to order $\varepsilon^2$, for different soliton speeds $\beta$. The analytical result obtained from \eqref{shift} and the numerical solutions show very good agreement, consistent with all our results being exact to the orders in $\varepsilon$ and $1/\lambda$ that we have expressed. The qualitative behavior of the soliton displacement as a function of the soliton speed $\beta$ and central wavenumber $k$ seen in Fig.~\eqref{figshift} can be understood in terms of the group velocity $\nu_k$ of the excitation packet. It is obvious that the interaction time between the soliton and the excitation depends on the difference of the soliton speed and the group velocity. Hence, the soliton displacement increases with increasing soliton speed $\beta$ and decreasing $k$.

To demonstrate the effect of the wave packet dressing contribution in $\psi_2$ on the soliton shift, we draw the analytical curve for the normalized soliton shift ignoring the contributions $N_2$ \eqref{n2} to the total number of excited atoms $N$ in Fig.~\ref{figshift}. Without the systematic inclusion of all second order pieces, the soliton shift is clearly underestimated, especially for low $k \lesssim 1$. This shows that the soliton back-reaction is in general a post-Bogoliubov effect, not only in the direct sense that the soliton motion zero mode is excited only at second order in $\varepsilon$, but also in that the back-reaction is sensitive to the general post-Bogoliubov dressing of the other excitations. The number of excited atoms $N_2$ exactly contributes $N/2$ to the total number of excited atoms $N$ for $k=0$. The ratio $N_2/N$ is a monotonically decreasing function of $k$, and in the opposite regime $k \gg 1$ one finds $N_2/N= 2 \mu/k^2+\mathcal{O}\left(k^{-3}\right)$; the excitation packet is dominated by the contribution $N_1$. 

%%%%%%%%%%%%%%%%%%%%%%%%%%%%%%%%%%%%%%%%%%%%%%
% Discussion
%%%%%%%%%%%%%%%%%%%%%%%%%%%%%%%%%%%%%%%%%%%%%%
\section{Discussion}
\label{discuss}
In conclusion, we have demonstrated that a soliton undergoes a finite shift as back-reaction to passing small-amplitude wave excitations. Within linear Bogoliubov theory we have shown that excitation pulses are shifted forward by passing a soliton. Solving the NLSE beyond linear order, we have then derived the soliton displacement back-reaction from exact conservation laws. The size of the back-reaction effect can qualitatively be understood in terms of the group velocity of the excitation and the soliton speed. The smaller the difference of the two velocities, the longer the interaction time between the excitation and the soliton, and the larger the back-reaction.

It is widely known that solitons are transparent to excitations, propagating through them without reflection; but our work here shows that this does not quite mean that solitons and excitations do not exchange momentum. What it means is that they have no net, permanent exchange of momentum. During the brief period when they overlap, however, they do exchange momentum temporarily. As an excitation passes through the soliton, it borrows some momentum from the soliton. It returns the full amount before leaving, but with the momentum it briefly borrows, it travels some extra distance. The soliton correspondingly loses momentum briefly, and its position shifts slightly until it gains the momentum back as the excitation departs. Because of the solitonÕs negative dynamical mass, this brief loss of momentum translates the soliton in the same direction as the excitation was advanced. The excitation is a trustworthy borrower of momentum, repaying its debt promptly and in full. The soliton is in turn a generous lender, since it charges no interest, but it is also a remarkably clever dealer, for after this peculiar transaction, both parties come out ahead. 

Our results shed some additional light on what thermal and quantum corrections to the simple GPE soliton must be like, because both thermal and quantum fluctuations can be represented as ensembles of wave packets like the ones we have discussed. On the one hand we can see that wave packets that are actually on top of the soliton tend to raise its $|\beta|$ temporarily. This means that if fluctuations really become strong, such that the soliton is continuously being passed by significant perturbations, then the darkest, narrowest solitons will indeed be smeared out, and become broader and grayer. On the other hand we can see that once a wave packet has passed the soliton, its lasting effect is only a translation. So in regimes with weaker fluctuations, we can expect to see something more like Brownian motion of the soliton, where it gets kicked back and forth at random intervals, while retaining its form. And we would expect the transition between these two regimes to be gradual rather than abrupt, since both effects are always present, and which dominates depends only on how many perturbation packets are typically present.

%%%%%%%%%%%%%%%%%%%%%%%%%%%%%%%%%%%%%%%%%%%%%%
%
%%%%%%%%%%%%%%%%%%%%%%%%%%%%%%%%%%%%%%%%%%%%%%
\appendix
\section{Group velocity to order $ \lambda^{-2}$}
\label{secondlambda}
An elegant way of rewriting the integrand of \eqref{exc} at second order is to use
\begin{equation}
\begin{split}
\bar{u}_{k+\zeta/\lambda} & =\bar{u}_k \exp\left[-\frac{ \zeta \bar{u}'_k}{\lambda \bar{u}_k}+\frac{\zeta^2}{ 2 \lambda^2} \left( \frac{\bar{u}''_k}{\bar{u}_k}- \left(\frac{\bar{u}'_k}{\bar{u}_k} \right)^2 \right)\right] \\
& \qquad \qquad \qquad \times\left\{1+\mathcal{O} \left(\frac{\zeta}{\lambda}\right)^3 \right\}
\end{split}
\end{equation}
as well as the analog for $ \bar{v}_k$. The Gaussian integrals can then be written and solved in a compact way and with the definition of the phase $ \phi \equiv k x \pm \theta_k/2-\Omega_k t$ we obtain outside the soliton range ($ \kappa |x| \gg1$):
\begin{multline}
\label{exc1dis}
\psi_{12} \! \equiv \frac{\psi_\gtrless}{| \psi_\gtrless|} \Bigg\{ \frac{\bar{u}_k e^{\mathrm{i} \phi}} {\sqrt{ \tilde{\lambda}^2_{\bar{u}}}} \exp\Bigg[-\frac{(z_\pm- \mathrm{i} \frac{\bar{u}_k'}{\bar{u}_k})^2}{2 \tilde{\lambda}^2_{\bar{u}}} \Bigg] \\
+( \bar{u}_k \leftrightarrow \bar{v}_k, \mathrm{c.c.}) \Bigg\}\, .
\end{multline}
The full expression of the excitation pulse at corresponding order is obtained by replacing $\bar{u}_k$ with $ \bar{v}_k$ in the first line and taking the complex conjugate. With the second subindex $i$ of $\psi_{1i}$ we indicate the order in $\lambda^{-1}$ we have kept. In \eqref{exc1dis} we have defined $\tilde{\lambda}^2(j_k)$ for an arbitrary function $j_k$ by
\begin{equation}
\label{gamma}
\begin{split}
\tilde{\lambda}^2_j &\equiv \lambda^2- \left(\frac{j_k''}{j_k}-\frac{j_k'^2}{j_k^2} \right) + \mathrm{i} \left( \Omega_k'' t \mp \frac{ \theta_k''}{2} \right) \; .
\end{split}
\end{equation}
The effect of $ \tilde{\lambda}^2$ in comparison to \eqref{exc11} can be most easily seen if we assume $ \varepsilon \ll1$, such that we can neglect terms of order $\varepsilon^2$ in $ |\psi_0+\varepsilon \psi_{12}|^2$. The envelope of the excitation is then determined by $ \psi_0 \psi_{12}^*+\psi_0^* \psi_{12}$. Since $\tilde{\lambda}^2$ is complex the envelope of $\psi_{12}$ is corrected at second order in $\lambda^{-2}$. Thus, the envelope is dispersive and the group velocity of the excitation pulse has increased to
\begin{equation}
\label{tildenu}
\tilde{\nu}_k \equiv \nu_k+\frac{(\bar{u}'_k+\bar{v}'_k)}{(\bar{u}_k+\bar{v}_k)} \frac{\Omega_k''}{\lambda^2}+\mathcal{O}(\lambda^{-4}) \; .
\end{equation}
It is important to notice that the change $\delta \nu_k \equiv \tilde{\nu}_k - \nu_k$ in the group velocity, an effect of order $\mathcal{O}(\lambda^{-2})$, is present for both constant and soliton background.

\bibliographystyle{apsrev}%abbrevmod

\begin{thebibliography}{26}
\expandafter\ifx\csname natexlab\endcsname\relax\def\natexlab#1{#1}\fi
\expandafter\ifx\csname bibnamefont\endcsname\relax
  \def\bibnamefont#1{#1}\fi
\expandafter\ifx\csname bibfnamefont\endcsname\relax
  \def\bibfnamefont#1{#1}\fi
\expandafter\ifx\csname citenamefont\endcsname\relax
  \def\citenamefont#1{#1}\fi
\expandafter\ifx\csname url\endcsname\relax
  \def\url#1{\texttt{#1}}\fi
\expandafter\ifx\csname urlprefix\endcsname\relax\def\urlprefix{URL }\fi
\providecommand{\bibinfo}[2]{#2}
\providecommand{\eprint}[2][]{\url{#2}}

\bibitem[{\citenamefont{Hawking}(1974)}]{Hawking1974}
\bibinfo{author}{\bibfnamefont{S.~W.} \bibnamefont{Hawking}},
  \bibinfo{journal}{Nature} \textbf{\bibinfo{volume}{248}}, \bibinfo{pages}{30}
  (\bibinfo{year}{1974}).

\bibitem[{\citenamefont{Hawking}(1975)}]{Hawking1975}
\bibinfo{author}{\bibfnamefont{S.~W.} \bibnamefont{Hawking}},
  \bibinfo{journal}{Communications in Mathematical Physics}
  \textbf{\bibinfo{volume}{43}}, \bibinfo{pages}{199} (\bibinfo{year}{1975}).

\bibitem[{\citenamefont{Unruh}(1976)}]{Unruh1976}
\bibinfo{author}{\bibfnamefont{W.~G.} \bibnamefont{Unruh}},
  \bibinfo{journal}{Physical Review D} \textbf{\bibinfo{volume}{14}},
  \bibinfo{pages}{870} (\bibinfo{year}{1976}).

\bibitem[{\citenamefont{Weller et~al.}(2008)\citenamefont{Weller, Ronzheimer,
  Gross, Esteve, Oberthaler, Frantzeskakis, Theocharis, and
  Kevrekidis}}]{Weller2008}
\bibinfo{author}{\bibfnamefont{A.}~\bibnamefont{Weller}},
  \bibinfo{author}{\bibfnamefont{J.~P.} \bibnamefont{Ronzheimer}},
  \bibinfo{author}{\bibfnamefont{C.}~\bibnamefont{Gross}},
  \bibinfo{author}{\bibfnamefont{J.}~\bibnamefont{Esteve}},
  \bibinfo{author}{\bibfnamefont{M.~K.} \bibnamefont{Oberthaler}},
  \bibinfo{author}{\bibfnamefont{D.~J.} \bibnamefont{Frantzeskakis}},
  \bibinfo{author}{\bibfnamefont{G.}~\bibnamefont{Theocharis}},
  \bibnamefont{and} \bibinfo{author}{\bibfnamefont{P.~G.}
  \bibnamefont{Kevrekidis}}, \bibinfo{journal}{Physical Review Letters}
  \textbf{\bibinfo{volume}{101}}, \bibinfo{pages}{130401}
  (\bibinfo{year}{2008}).

\bibitem[{\citenamefont{Stellmer et~al.}(2008)\citenamefont{Stellmer, Becker,
  Soltan-Panahi, Richter, Dorscher, Baumert, Kronjager, Bongs, and
  Sengstock}}]{Stellmer2008}
\bibinfo{author}{\bibfnamefont{S.}~\bibnamefont{Stellmer}},
  \bibinfo{author}{\bibfnamefont{C.}~\bibnamefont{Becker}},
  \bibinfo{author}{\bibfnamefont{P.}~\bibnamefont{Soltan-Panahi}},
  \bibinfo{author}{\bibfnamefont{E.~M.} \bibnamefont{Richter}},
  \bibinfo{author}{\bibfnamefont{S.}~\bibnamefont{Dorscher}},
  \bibinfo{author}{\bibfnamefont{M.}~\bibnamefont{Baumert}},
  \bibinfo{author}{\bibfnamefont{J.}~\bibnamefont{Kronjager}},
  \bibinfo{author}{\bibfnamefont{K.}~\bibnamefont{Bongs}}, \bibnamefont{and}
  \bibinfo{author}{\bibfnamefont{K.}~\bibnamefont{Sengstock}},
  \bibinfo{journal}{Physical Review Letters} \textbf{\bibinfo{volume}{101}},
  \bibinfo{pages}{120406} (\bibinfo{year}{2008}).

\bibitem[{\citenamefont{Becker et~al.}(2008)\citenamefont{Becker, Stellmer,
  Soltan-Panahi, Dorscher, Baumert, Richter, Kronjager, Bongs, and
  Sengstock}}]{Becker2008}
\bibinfo{author}{\bibfnamefont{C.}~\bibnamefont{Becker}},
  \bibinfo{author}{\bibfnamefont{S.}~\bibnamefont{Stellmer}},
  \bibinfo{author}{\bibfnamefont{P.}~\bibnamefont{Soltan-Panahi}},
  \bibinfo{author}{\bibfnamefont{S.}~\bibnamefont{Dorscher}},
  \bibinfo{author}{\bibfnamefont{M.}~\bibnamefont{Baumert}},
  \bibinfo{author}{\bibfnamefont{E.~M.} \bibnamefont{Richter}},
  \bibinfo{author}{\bibfnamefont{J.}~\bibnamefont{Kronjager}},
  \bibinfo{author}{\bibfnamefont{K.}~\bibnamefont{Bongs}}, \bibnamefont{and}
  \bibinfo{author}{\bibfnamefont{K.}~\bibnamefont{Sengstock}},
  \bibinfo{journal}{Nature Physics} \textbf{\bibinfo{volume}{4}},
  \bibinfo{pages}{496} (\bibinfo{year}{2008}).

\bibitem[{\citenamefont{Tsuzuki}(1971)}]{Tsuzuki1971}
\bibinfo{author}{\bibfnamefont{T.}~\bibnamefont{Tsuzuki}},
  \bibinfo{journal}{Journal of Low Temperature Physics}
  \textbf{\bibinfo{volume}{4}}, \bibinfo{pages}{441} (\bibinfo{year}{1971}).

\bibitem[{\citenamefont{Gross}(1961)}]{Gross1961}
\bibinfo{author}{\bibfnamefont{E.~P.} \bibnamefont{Gross}},
  \bibinfo{journal}{Nuovo Cimento} \textbf{\bibinfo{volume}{20}},
  \bibinfo{pages}{454} (\bibinfo{year}{1961}).

\bibitem[{\citenamefont{Pitaevskii}(1961)}]{Pitaevskii1961}
\bibinfo{author}{\bibfnamefont{L.~P.} \bibnamefont{Pitaevskii}},
  \bibinfo{journal}{Soviet Physics JETP} \textbf{\bibinfo{volume}{13}},
  \bibinfo{pages}{451} (\bibinfo{year}{1961}).

\bibitem[{\citenamefont{Walczak and Anglin}(2011)}]{Walczak2011}
\bibinfo{author}{\bibfnamefont{P.~B.} \bibnamefont{Walczak}} \bibnamefont{and}
  \bibinfo{author}{\bibfnamefont{J.~R.} \bibnamefont{Anglin}},
  \bibinfo{journal}{Physical Review A} \textbf{\bibinfo{volume}{84}},
  \bibinfo{pages}{013611} (\bibinfo{year}{2011}).

\bibitem[{\citenamefont{Busch and Anglin}(2000)}]{Busch2000}
\bibinfo{author}{\bibfnamefont{T.}~\bibnamefont{Busch}} \bibnamefont{and}
  \bibinfo{author}{\bibfnamefont{J.~R.} \bibnamefont{Anglin}},
  \bibinfo{journal}{Physical Review Letters} \textbf{\bibinfo{volume}{84}},
  \bibinfo{pages}{2298} (\bibinfo{year}{2000}).

\bibitem[{\citenamefont{Muryshev et~al.}(2002)\citenamefont{Muryshev,
  Shlyapnikov, Ertmer, Sengstock, and Lewenstein}}]{Muryshev2002}
\bibinfo{author}{\bibfnamefont{A.}~\bibnamefont{Muryshev}},
  \bibinfo{author}{\bibfnamefont{G.~V.} \bibnamefont{Shlyapnikov}},
  \bibinfo{author}{\bibfnamefont{W.}~\bibnamefont{Ertmer}},
  \bibinfo{author}{\bibfnamefont{K.}~\bibnamefont{Sengstock}},
  \bibnamefont{and}
  \bibinfo{author}{\bibfnamefont{M.}~\bibnamefont{Lewenstein}},
  \bibinfo{journal}{Physical Review Letters} \textbf{\bibinfo{volume}{89}},
  \bibinfo{pages}{110401} (\bibinfo{year}{2002}).

\bibitem[{\citenamefont{Mazets et~al.}(2008)\citenamefont{Mazets, Schumm, and
  Schmiedmayer}}]{Mazets2008}
\bibinfo{author}{\bibfnamefont{I.~E.} \bibnamefont{Mazets}},
  \bibinfo{author}{\bibfnamefont{T.}~\bibnamefont{Schumm}}, \bibnamefont{and}
  \bibinfo{author}{\bibfnamefont{J.}~\bibnamefont{Schmiedmayer}},
  \bibinfo{journal}{Physical Review Letters} \textbf{\bibinfo{volume}{100}},
  \bibinfo{pages}{210403} (\bibinfo{year}{2008}).

\bibitem[{\citenamefont{Martin and
  Ruostekoski}(2010{\natexlab{a}})}]{Martin2010}
\bibinfo{author}{\bibfnamefont{A.~D.} \bibnamefont{Martin}} \bibnamefont{and}
  \bibinfo{author}{\bibfnamefont{J.}~\bibnamefont{Ruostekoski}},
  \bibinfo{journal}{Physical Review Letters} \textbf{\bibinfo{volume}{104}},
  \bibinfo{pages}{194102} (\bibinfo{year}{2010}{\natexlab{a}}).

\bibitem[{\citenamefont{Martin and
  Ruostekoski}(2010{\natexlab{b}})}]{Martin2010a}
\bibinfo{author}{\bibfnamefont{A.~D.} \bibnamefont{Martin}} \bibnamefont{and}
  \bibinfo{author}{\bibfnamefont{J.}~\bibnamefont{Ruostekoski}},
  \bibinfo{journal}{New Journal Of Physics} \textbf{\bibinfo{volume}{12}},
  \bibinfo{pages}{055018} (\bibinfo{year}{2010}{\natexlab{b}}).

\bibitem[{\citenamefont{Steel et~al.}(1998)\citenamefont{Steel, Olsen, Plimak,
  Drummond, Tan, Collett, Walls, and Graham}}]{Steel1998}
\bibinfo{author}{\bibfnamefont{M.~J.} \bibnamefont{Steel}},
  \bibinfo{author}{\bibfnamefont{M.~K.} \bibnamefont{Olsen}},
  \bibinfo{author}{\bibfnamefont{L.~I.} \bibnamefont{Plimak}},
  \bibinfo{author}{\bibfnamefont{P.~D.} \bibnamefont{Drummond}},
  \bibinfo{author}{\bibfnamefont{S.~M.} \bibnamefont{Tan}},
  \bibinfo{author}{\bibfnamefont{M.~J.} \bibnamefont{Collett}},
  \bibinfo{author}{\bibfnamefont{D.~F.} \bibnamefont{Walls}}, \bibnamefont{and}
  \bibinfo{author}{\bibfnamefont{R.}~\bibnamefont{Graham}},
  \bibinfo{journal}{Physical Review A} \textbf{\bibinfo{volume}{58}},
  \bibinfo{pages}{4824} (\bibinfo{year}{1998}).

\bibitem[{\citenamefont{Cockburn et~al.}(2010)\citenamefont{Cockburn,
  Nistazakis, Horikis, Kevrekidis, Proukakis, and
  Frantzeskakis}}]{Cockburn2010}
\bibinfo{author}{\bibfnamefont{S.~P.} \bibnamefont{Cockburn}},
  \bibinfo{author}{\bibfnamefont{H.~E.} \bibnamefont{Nistazakis}},
  \bibinfo{author}{\bibfnamefont{T.~P.} \bibnamefont{Horikis}},
  \bibinfo{author}{\bibfnamefont{P.~G.} \bibnamefont{Kevrekidis}},
  \bibinfo{author}{\bibfnamefont{N.~P.} \bibnamefont{Proukakis}},
  \bibnamefont{and} \bibinfo{author}{\bibfnamefont{D.~J.}
  \bibnamefont{Frantzeskakis}}, \bibinfo{journal}{Physical Review Letters}
  \textbf{\bibinfo{volume}{104}}, \bibinfo{pages}{174101}
  (\bibinfo{year}{2010}).

\bibitem[{\citenamefont{Stoof}(1999)}]{Stoof1999}
\bibinfo{author}{\bibfnamefont{H.~T.~C.} \bibnamefont{Stoof}},
  \bibinfo{journal}{Journal Of Low Temperature Physics}
  \textbf{\bibinfo{volume}{114}}, \bibinfo{pages}{11} (\bibinfo{year}{1999}).

\bibitem[{\citenamefont{Mishmash and Carr}(2009)}]{Mishmash2009}
\bibinfo{author}{\bibfnamefont{R.~V.} \bibnamefont{Mishmash}} \bibnamefont{and}
  \bibinfo{author}{\bibfnamefont{L.~D.} \bibnamefont{Carr}},
  \bibinfo{journal}{Physical Review Letters} \textbf{\bibinfo{volume}{103}},
  \bibinfo{pages}{140403} (\bibinfo{year}{2009}).

\bibitem[{\citenamefont{Mishmash et~al.}(2009)\citenamefont{Mishmash, Danshita,
  Clark, and Carr}}]{Mishmash2009a}
\bibinfo{author}{\bibfnamefont{R.~V.} \bibnamefont{Mishmash}},
  \bibinfo{author}{\bibfnamefont{I.}~\bibnamefont{Danshita}},
  \bibinfo{author}{\bibfnamefont{C.~W.} \bibnamefont{Clark}}, \bibnamefont{and}
  \bibinfo{author}{\bibfnamefont{L.~D.} \bibnamefont{Carr}},
  \bibinfo{journal}{Physical Review A} \textbf{\bibinfo{volume}{80}},
  \bibinfo{pages}{053612} (\bibinfo{year}{2009}).

\bibitem[{\citenamefont{Pitaevskii and Stringari}(2003)}]{Pitaevskii2003}
\bibinfo{author}{\bibfnamefont{L.}~\bibnamefont{Pitaevskii}} \bibnamefont{and}
  \bibinfo{author}{\bibfnamefont{S.}~\bibnamefont{Stringari}},
  \emph{\bibinfo{title}{Bose-Einstein Condensation}}, no. \bibinfo{number}{116}
  in \bibinfo{series}{International Series of Monographs on Physics}
  (\bibinfo{publisher}{Oxford University Press}, \bibinfo{address}{New York},
  \bibinfo{year}{2003}).

\bibitem[{\citenamefont{Fetter}(1972)}]{Fetter1972}
\bibinfo{author}{\bibfnamefont{A.~L.} \bibnamefont{Fetter}},
  \bibinfo{journal}{Annals Of Physics} \textbf{\bibinfo{volume}{70}},
  \bibinfo{pages}{67} (\bibinfo{year}{1972}).

\bibitem[{\citenamefont{Pethick and Smith}(2008)}]{Pethick2008}
\bibinfo{author}{\bibfnamefont{C.}~\bibnamefont{Pethick}} \bibnamefont{and}
  \bibinfo{author}{\bibfnamefont{H.}~\bibnamefont{Smith}},
  \emph{\bibinfo{title}{Bose-Einstein Condensation in Dilute Gases}}
  (\bibinfo{publisher}{Cambridge University Press, Cambridge},
  \bibinfo{year}{2008}), \bibinfo{edition}{2nd} ed.

\bibitem[{\citenamefont{Negretti et~al.}(2008)\citenamefont{Negretti, Henkel,
  and M\o~lmer}}]{Negretti2008a}
\bibinfo{author}{\bibfnamefont{A.}~\bibnamefont{Negretti}},
  \bibinfo{author}{\bibfnamefont{C.}~\bibnamefont{Henkel}}, \bibnamefont{and}
  \bibinfo{author}{\bibfnamefont{K.}~\bibnamefont{M\o~lmer}},
  \bibinfo{journal}{Physical Review A} \textbf{\bibinfo{volume}{78}},
  \bibinfo{pages}{023630} (\bibinfo{year}{2008}).

\bibitem[{\citenamefont{Weideman and Herbst}(1986)}]{Weideman1986}
\bibinfo{author}{\bibfnamefont{J.~A.~C.} \bibnamefont{Weideman}}
  \bibnamefont{and} \bibinfo{author}{\bibfnamefont{B.~M.}
  \bibnamefont{Herbst}}, \bibinfo{journal}{Siam Journal On Numerical Analysis}
  \textbf{\bibinfo{volume}{23}}, \bibinfo{pages}{485} (\bibinfo{year}{1986}).

\bibitem[{\citenamefont{Stringari}(1996)}]{Stringari1996}
\bibinfo{author}{\bibfnamefont{S.}~\bibnamefont{Stringari}},
  \bibinfo{journal}{Physical Review Letters} \textbf{\bibinfo{volume}{77}},
  \bibinfo{pages}{2360} (\bibinfo{year}{1996}).

\end{thebibliography}

\end{document}